\documentclass[mathleft
]{an}
\usepackage{graphicx}
\usepackage{times}
\begin{document}

\Pagespan{789}{}
\Yearpublication{2006}%
\Yearsubmission{2005}%
\Month{11}%
\Volume{999}%
\Issue{88}%

\title{Models for the X-ray spectra and variability of
luminous accreting black holes}

\author{J. Malzac\inst{1}
\fnmsep\thanks{Corresponding author:\email{malzac@cesr.fr}\newline}
\and  
A. Merloni\inst{2}
\and
T. Suebsuwong\inst{1}}

\titlerunning{Luminous accreting black holes}
\authorrunning{Malzac et al.}
\institute{
CESR (CNRS/UPS/OMP), 9 Avenue du colonel Roche, BP4346  31028 Toulouse, Cedex 4, France 
\and 
Max-Planck-Institut f\"ur Astrophysik,Karl-Schwarzschild-Strasse 1, D-85741, Garching, Germany
}
\received{30 May 2005}
\accepted{11 Nov 2005}
\publonline{later}

\keywords{accretion, accretion disks -- black hole physics -- radiative transfer --  relativity}

\abstract{%
 The X-ray spectra of luminous Seyfert 1 galaxies often appear to be reflection dominated.
In a number of Narrow Line Seyfert 1 (NLS1) galaxies and galactic
black holes in the very high state, the variability of the continuum and of the
iron line are decoupled, the reflected component being often much less variable
than the continuum.
These properties have been interpreted as effects of gravitational light bending.
In this framework, we present detailed Monte-Carlo simulations of the
reflection continuum in the Kerr metric. These calculations confirrm that the
spectra and variability behaviour of these sources can be reproduced by the
light bending model.
As an alternative to the light bending model, we show that similar observational
properties are expected from radiation pressure dominated discs subject
to violent clumping instabilities and, as a result, have a highly inhomogeneous
two-phase structure. In this model, most of the observed spectral and variability
features originate from the complex geometrical structure of the 
inner regions of near-Eddington accretion flows and are therefore a signature
of accretion physics rather than general relativity.}

\maketitle

\section{Introduction}
The spectra of  luminous Seyfert galaxies are very well described by photoionized and strongly relativistically blurred reflection models (Fabian et al. 2004, 2005; Crummy et al. 2006; Porquet  2006).
 In these sources the primary continuum often appears to be strongly suppressed. In a number of Narrow Line Seyfert 1 (NLS1) galaxies and galactic black holes in the very high state, the variabilities of the continuum and of the iron line are decoupled, in apparent contradiction with the predictions of simple disc reflection models (see, e.g.  Miniutti, Fabian \& Miller 2004; Fabian et al. 2004).  In particular, the monitoring of Seyfert galaxies indicates that the reflection
  flux  can be weakly variable when the primary emission
  changes dramatically (Papadakis et al. 2002). Moreover in at least two AGN, MGC 6-30-15 (Miniutti et al 2003)
   and NGC4051 (Ponti et al. 2006)  and one X-ray binary
  (XTE J1650-500, Rossi et al. 2005)  the reflection flux is correlated to the primary
  emission at low fluxes and saturates at higher fluxes. In this paper we discuss two possible models for these puzzling properties.  In the light bending model of Miniutti \& Fabian (2004) they are interpreted in terms of general relativistic effects, while, in the clumpy disc model of Merloni et al. (2006),  these properties would be the result of the complex structure of near Eddington accretion flows. 
  
\section{Light bending model}

\begin{figure}
   \centering
   \includegraphics[width=8cm]{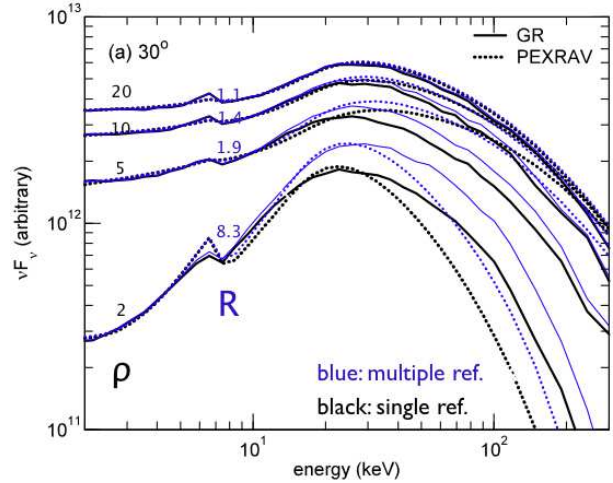} 
   \caption{Light bending model spectra (light blue lines) for a ring height of 2 $R_{g}$ and various radii 
   $\rho =$ 2, 5, 10 and 20 $R_g$, from bottom to top respectively. The thick black lines show the spectra obtained when multiple reflections, due to radiation returning to the disc, are neglected. The effects of multiple reflections are very significant above 10 keV .  The dotted curves show the Newtonian model (PEXRAV) providing the best fit of the ligth bending spectra in the 2-30 keV band. The reflected spectra are for neutral material and standard abundances.The primary emission consists in an e-folded power law with photon index $\Gamma=2$ and $E_c=200$ keV. The inclination angle is $30^o$. The spectra are reflection dominated when the ring radius is low,  as can be seen from the reflection coefficients  $R>$ 1 shown on the figure.}
   \label{fig:specti}
\end{figure}

In this model, the active coronal region(s) illuminating the disc are idealised as a ring source at some height above, and corotating with the accretion disc. When the source is close enough to the black hole, the primary component is strongly suppressed leading to reflection dominated spectra. Moreover, as shown by Miniutti \& Fabian (2004), fluctuations in the height of the source can lead to strong variability in the primary component with little variability in the reflected flux, as observed.  
These results were recently confirmed by Suebsuwong et al. (2006) which present calculations of the light bending model spectra. These calculations improve
upon the previous works by fully computing broad band angle dependent reflection spectra and primary emission as a function of  ring radius, $\rho$, and height $h$, including the effects of multiple reflection.
Fig.~\ref{fig:specti} show some sample spectra obtained when varying the ring radius while keeping a constant height above the disc ($h=2$ $R_g$) . These spectra show that the smaller the radius, the stronger the reflection component. Indeed when the primary X-ray source is located at less than a few $R_g$ from the black hole horizon, the light bending effects are very strong and tend to beam the primary radiation toward the disc, leading to strongly reflection dominated spectra. In these extreme cases the irradiation is concentrated 
in the central part of the disc. Then the same light bending effects make it difficult for the reflected photons to escape. A significant fraction of them returns to the disc where they can be reflected again and so on. 
The reflection spectrum is then made up of the sum of the multiply reflected spectra, which make it significantly different above 10 keV from what is expected from single reflection models (as shown in Fig.~\ref{fig:specti})
These effects should be taken into account when fitting the spectra of
  extreme sources such as MGC 6-30-15
  with broad band instruments such as BeppoSAX, INTEGRAL
  or Suzaku.
  
\begin{figure}
   \centering
   \includegraphics[width=8cm]{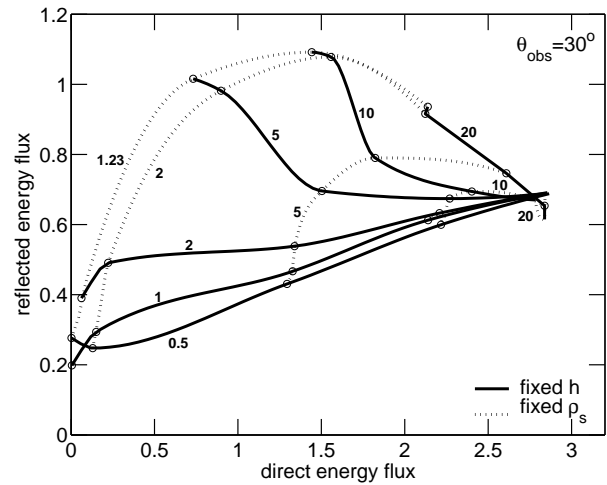} 
   \caption{The total reflected energy flux as a function of the total direct energy flux,
both are in the energy range 1-30 keV, for an observation inclination $30^o$
 and  different source height and radius.
 The thin lines represent the fixed source radius and the thick lines indicate the same
  source height. These curves are spline interpolations between the results of 
  the light bending model spectra simulations lying at their intersections. }
   \label{fig:flpvsflref}
\end{figure}

Figure~\ref{fig:flpvsflref} shows the dependence of the observed reflected and primary luminosity upon the height $h$ and radius $\rho$ of the ring source. Fluctuations of the geometrical parameters of the source lead to changes in the reflected and primary flux that are not always correlated.  Indeed,  as can be seen on this figure, depending on the parameter regime, the reflected and primary components can also be anti-correlated or nearly independent. 
When the source height changes at constant radius and as long as $\rho_{s}< 5 R_g$,
 its track in this plane can be described according to three regimes:
i) at low fluxes (or low source height) the reflected and primary flux are correlated,
ii) at higher fluxes the reflection saturates at an almost constant value
while the primary can change by a factor larger than 2,
iii)  at even higher fluxes the reflection component is weakly anti-correlated
with the primary emission.
This behaviour is described in more details by Miniutti and Fabian (2004).
 As shown by these authors many properties of the variability of Seyfert galaxies
  and black hole binaries can be understood in terms of fluctuations of the source height.
  In particular the curious behaviour of  MGC 6-30-15, NGC4051, XTE J1650-500, where the reflection flux is correlated to the primary
  emission at low fluxes and saturates at higher fluxes, is in qualitative agreement with
   the predictions of this model.
Fig~\ref{fig:flpvsflref} enables us to investigate further the model parameter space.
If the radius is larger than $\sim  5 R_g$ the variability induced
 by change in the height is much too weak ($< 2$) to account for the variability observed in most accreting black holes.
Let now consider the effects of changes in the source radius at constant height.
At small source heights ($h<5 R_g$), the overall trend is that the reflection and
 primary emission are weakly correlated: the reflected flux changes by at most 50 \%
 when the primary flux increases by more than one order of magnitude which might be in qualitative agreement
 with some observations but is inconsistent with the strong non-linear
 correlation observed  for instance in the low state of NGC4051.
At higher source heights, the reflected and primary fluxes become
 anti-correlated, which is not observed.
The slope of the anti-correlation increases with $h$.
At $h\sim 10 R_g$ we could observe large variations of the reflection
  component at constant primary flux.
These results suggest that if the light bending model is to be the correct
  interpretation of the observations,
  the driver of the variability should be $h$
  while the source ring radius has to be nearly constant and small ($< 5 R_g$).

\begin{figure}
 \centering
   \includegraphics[width=8cm]{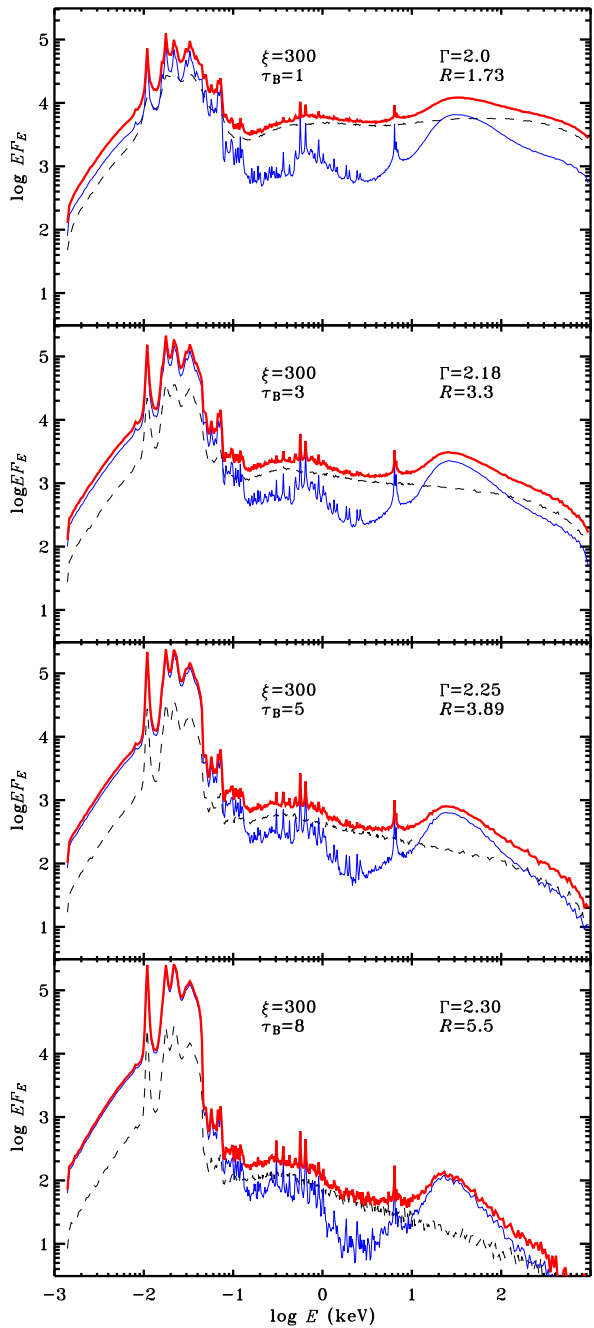} 
   \caption{Inhomogenous disc model: angle averaged spectra for $\tau_{\rm B}$ ranging from 1 to 8 as indicated. The best fit parameter $R$ and $\Gamma$  obtained when these spectra are fitted with PEXRAV in the 2-30 keV range are shown as well. The other fixed  model parameters are the vertical Thomson optical of the disc $\tau_{\rm T}=1$, the ionisation parameter of the of the cold clumps $\xi=300$, and the size of the regions where the plasma is heated $h=0.1H$ (see Merloni et al. 2006 for details). }
   \label{fig:spectaubw}
\end{figure}

\begin{figure}
 \centering
   \includegraphics[angle=-90,width=8cm]{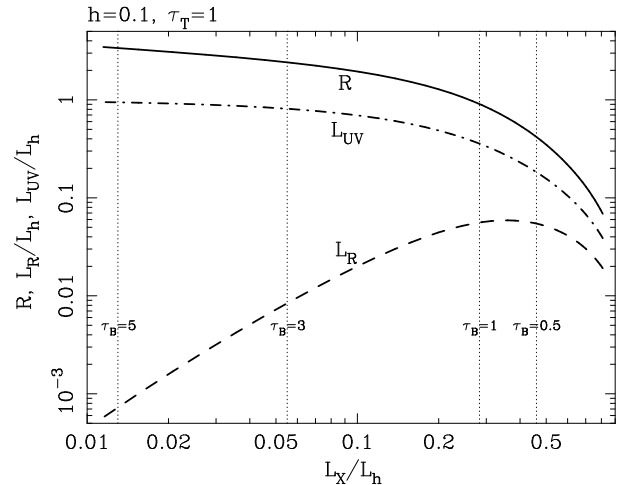} 
   \caption{Inhomogeneous disc: the relative intensity of the reflection component ($R$,
   solid line), of the reprocessed UV ($L_{\rm UV}$, dot-dashed line) and  reflected
   ($L_{\rm R}$, dashed line) luminosities are plotted as
  functions of the emergent X-ray (Comptonised power-law)
 luminosity above 1 keV ($L_{\rm X}$) for a varying
   cloud optical depth $\tau_{\rm B}$.
   All luminosities are renormalized to the total heating rate $L_h$ (see Merloni et al. 2006 for details).
   The vertical dotted lines mark the X-ray luminosities
   corresponding to $\tau_{\rm B}$=0.5,1,3,5 }
   \label{fig:h01_1_alR04}
\end{figure}

\section{Inhomogeneous accretion flows}

Black Hole systems, either in X-ray
binaries or in active galactic nuclei,
when shining at luminosities close to
the Eddington limit are thought to be powered by accretion  \linebreak 
through geometrically thin, optically thick discs.
Nonetheless, according to the
  standard accretion disc solutions, such
  highly luminous discs should be
  radiation pressure dominated and therefore
  unstable to perturbations of both mass flow (Lightman \& Eardley
  1974) and heating rates (Shakura \& Sunyaev 1973).
Thus, it is not clear yet to what extent this standard solutions
  represent a realistic description of the observed systems.
In recent years, both analytic works
(Blaes and Socrates 2001, 2003)
and simulations (Turner, Stone and Sano 2002, Turner et al. 2003,
Turner 2004) have shown that magnetized,
radiation pressure dominated accretion discs may be in fact
subject to violent clumping instabilities if magnetic pressure exceeds
  gas pressure and photons are able to diffuse from compressed regions.
Large density variations may also be caused by photon bubble instabilities
(Gammie 1998), which may develop into a series of shocks propagating
  through the plasma (Begelman 2001,2002; Turner at al. 2005).
These instabilities may in turn have profound
effects not only on the nature of the cooling mechanism of luminous
  discs, but also on their observational appearance (see e.g.  Ballantyne  et al. 2004).

Recently Merloni et al. (2006, hereafter M06) explored in detail the consequences of the hypothesis
that accretion flows close to the Eddington rate are indeed 
inhomogeneous both in its density and heating structure.
In this model we assume that the disc consist in a 
inhomogeneously heated plasma pervaded by small cold dense clumps.
M06  present a detailed study of their expected X-ray
spectral and variability, taking into account the radiative coupling between the two phases.
 The spectrum emerging from this inhomogeneous, two-phase, 
 clumpy flow is calculated by coupling the non-linear Monte Carlo code of Malzac \& Jourdain (2000), accounting for the Compton equilibrium between the cold and the hot phase, with the X-ray
ionized reflection code of Ross \& Fabian (1993, 2005), that accurately
computes the reprocessed radiation in the cold phase. 
 A crucial factor in determining the broad band spectral properties of an inhomogeneous flow
is the amount of cold clouds pervading the hot
plasma. In the limit of small scale optically thick clumps, this is quantified by the cloud optical depth $\tau_{\rm B}=nSH$, where $H$ is the disc scale height ,  $n$ is the cold clouds number density and $S$ their average geometrical cross section.
 Figure~\ref{fig:spectaubw} shows a sequence of spectra obtained by varying $\tau_{\rm B}$.  As the cloud covering fraction (i.e. $\tau_{\rm B}$) increases, the primary spectrum becomes softer, because of the enhanced cooling in the hot phase, and at the same time the reflection/reprocessing features become more and more prominent. Reflection dominated spectra are achieved for $\tau_{\rm B}>1$.
 As these spectra are supposed to be formed in the central part of the accretion flow, one naturally expect some relativistic blurring that is  not taken into account in the spectra of Fig.~\ref{fig:spectaubw}.
  
 M06 also derived analytical formulae to estimate the luminosity of the different spectral components of the radiation emerging from the inhomogeneous accretion flow.
 Fig.~\ref{fig:h01_1_alR04}  shows the reflected luminosity,
the soft reprocessed luminosity and the reflection fraction as a
function of the X-ray (Comptonized) luminosity. The fact that the reflected luminosity has a maximum, implies that large variations of the emergent X-ray luminosity, $L_{\rm X}$,
associated with changes in the cold clump integrated optical depth
correspond to only modest variations of the reflection component, at least as
long as $L_{\rm X}/L_{\rm h} \ga 0.1$.
On the other hand, for low values of the Comptonized X-ray luminosity,
the reflected luminosity correlates with $L_{\rm X}$,
  while at high $L_{\rm X}$, the trend is the opposite.
This global behaviour of $L_{\rm R}$ vs. $L_{\rm X}$ 
is strikingly similar  to that of the light bending model (see Fig~\ref{fig:flpvsflref}) and reproduces  the variability properties of the
continuum and of the iron line in a
number of Narrow Line Seyfert 1 (NLS1) galaxies and galactic black 
holes in the very high state .

\section{Light bending or inhomogenous accretion ?}  

These similarities between the light bending model and inhomogeneous disc models
lends itself to a simple geometrical
explanation. In the more general framework of two-phase models for the
X-ray
spectra of accreting black holes, the main spectral and variability
properties are determined by the geometry
(and on the topology) of the two phases and, in particular, on the sky covering
fraction of the cold phase as seen by the hot, Comptonising medium.
Reflection dominated spectra are expected when the cold phase
intercepts most of the photons coming from the hot phase. This, in the
light bending model, is achieved via general relativistic 
effects, while in the inhomogeneous disc model it is a result of the clumpy
 and inhomogeneous nature of the inner disc.

In principle, the relativistic blurring induced by the differential rotation of
the inner disc should always be taken into account when fitting
observed spectra. In the original light-bending model, where the
illuminating source is a \linebreak point-like source above a standard
geometrically thin disc, the ratio of the reflected component to the
power-law continuum is determined by the same effects that determine
 the shape of the relativistic lines, while if the disc is truly
 inhomogeneous, the two effects can be decoupled. Therefore,
 simultaneous spectroscopic studies of relativistically blurred
 emission lines and of the broad band continuum and variability could
 be effectively used to discriminate between a pure light bending
 model and a clumpy disc. Detailed predictions for the latter,
 however, require the combination of sophisticated MHD and radiative
 transfer simulations.



\end{document}